\documentclass[5p]{elsarticle}

\usepackage{listings}

\journal{Journal of \LaTeX\ Templates}

\biboptions{authoryear}

\begin{document}

\begin{frontmatter}

\title{ESASky : A Science-Driven Discovery Portal for Space-Based Astronomy Missions}

\author[1]{Fabrizio Giordano\fnref{myfootnote}}
\fntext[myfootnote]{fabrizio.giordano@esa.int}
\author[1]{Elena Racero}
\author[2]{Henrik Norman}
\author[3]{Ricardo Vall\'es}
\author[4]{Bruno Mer\'{\i}n}
\author[6]{Deborah Baines}
\author[5]{Marcos L\'opez-Caniego}
\author[7]{Bel\'en L\'opez Mart\'{\i}}
\author[8]{Pilar de Teodoro}
\author[6]{Jes\'us Salgado}
\author[8]{Maria Henar Sarmiento}
\author[7]{Raul Guti\'errez-S\'anchez}
\author[8]{Roberto Prieto} 
\author[8]{Alejandro Lorca}
\author[4]{Sara Alberola}
\author[7]{Ivan Valtchanov}
\author[9]{Guido de Marchi}
\author[4]{Rub\'en \'Alvarez} 
\author[4]{Christophe Arviset}

\address[1]{ESAC Science Data Centre, SERCO for ESA-ESAC, Spain}
\address[2]{Winter Way AB for ESA-ESAC, Sweden}
\address[3]{RHEA for ESA-ESAC, Spain}
\address[4]{ESAC Science Data Centre, European Space Agency (ESA), Spain}
\address[5]{ESAC Science Data Centre, AURORA for ESA-ESAC, Spain}
\address[6]{ESAC Science Data Centre, QUASAR Science Resources for ESA-ESAC, Spain}
\address[7]{Telespazio-Vega UK Ltd. for ESA-ESAC, Spain}
\address[8]{AURORA for ESA-ESAC, Spain}
\address[9]{ESAC Science Data Centre, ESA-ESTEC, The Netherlands}

\begin{abstract}

In the era of "big data" and with the advent of web 2.0 technologies, ESASky (http://sky.esa.int) aims at providing a modern and visual way to access astronomical science-ready data products and metadata. The main goal of the application is to simplify the interaction between the scientific community and the ever-growing amount of data collected over the past decades from the most important astronomy missions. 

The ESASky concept is to offer a complementary scientific application to the more-traditional table-oriented exploitation of astronomical data, by allowing a more natural and visual approach and enabling the exploration of astronomical objects across the entire electromagnetic spectrum. 
To fulfill this goal, ESASky provides a multiwavelength interface to a set of astronomy data from an increasing number of missions and surveys, with the intention of becoming the single-point of entry to perform visual analysis and cross-matching among different energy ranges. A lot of effort has been invested on the design of a user-friendly, responsive Graphical User Interface (GUI) by the definition and optimisation of algorithms running behind each visual feature offered.

In this contribution, we describe in detail the design and solutions adopted for the technical challenges arising during the development. We present the data services and features implemented in the latest version of ESASky (v2.1), including  a Mission Planning Tool to support current James Webb Space Telescope (JWST) planning, the possibility to search for observations of Solar System Objects (planets, comets and moons) taken by astronomy missions, the integration of the SAO/NASA Astrophysics Data System (ADS) publication system in the ESASky GUI, and the retrieval of metadata and data products available within a specific region of the sky.

\end{abstract}

\vspace{-0.5cm}

\end{frontmatter}

\section{Introduction}

The fast-growing volume of astronomical datasets and their increasing complexity makes it difficult for science data centres to serve the needs of both experts, with in-depth knowledge of the data, and of non-experts, either scientists from other fields of astronomy or even the general public. These datasets have been produced with a wide variety of instruments, at different wavelengths and resolutions, and with both ground-based and space-borne observatories. The tools needed to explore and analyse the data are, in general, complex and specific to an instrument or observatory, which makes it very difficult for scientists trying to perform multiwavelength analyses to exploit the large amount of already available archival data. This problem has been addressed to some extent by the International Virtual Observatory Alliance, introducing a set of standards and protocols that data centres can use to serve their data in a more efficient way. However, this is not enough to serve the needs of the astronomical community, which is eager to explore and exploit very complex datasets. 

With the goals of easing the tasks of retrieving and analysing multiwavelength astronomical data, the ESAC Science Data Centre (ESDC\footnote{http://www.cosmos.esa.int/web/esdc}; \citealt{CA:2015}) based at the European Space Astronomy Centre (ESAC) of the European Space Agency (ESA) near Madrid, Spain, has developed ESASky,\footnote{http://sky.esa.int} a science-driven discovery portal providing simplified access to astronomical data (\citealt{BM:2015}, \citealt{BLM:2016}, \citealt{DB:2017}).

ESASky enables quick and efficient searches of science-ready data products, namely images, catalogues and spectra, not only from ESA missions such as Planck, Herschel, Gaia, HST, XMM-Newton or INTEGRAL, but also from many other projects such as Chandra and Suzaku. All imaging observations from a given mission can be quickly visualised and compared with those of other missions thanks to a number of all-sky maps built from the actual observations from these missions. These maps are based on the Hierarchical Progressive Surveys (HiPS) technology \citep{PF:2015}, developed at the Centre de Donn\'ees Astronomiques de Strasbourg (CDS), in France.

In addition, ESASky allows users to simultaneously overlay the footprints of different datasets, providing a unique global view of the available multiwavelength observations of a given source or region of interest. Users can now visually compare the coverage of all observations available, filter and preview the data, and download the products of their interest, without having to visit every mission archive. The application also allows quick searches for images with Solar System Objects (SSO) in their Field of View (FoV). 

Moreover, the latest version of ESASky (v2.1), released in February 2018, gives the user the opportunity to search for all sources with scientific publications from the SAO/NASA Astrophysics Data System (ADS; \citealt{KM:2000}) in a given FoV. It also enables users to randomly browse predefined lists of targets, lists provided by the users or collected from publications from the ADS (stored in the SIMBAD database from the CDS). 

All these services and functionalities are made possible thanks to the combination of a user-friendly graphical user interface front-end with a powerful and efficient back-end engine. The purpose of this paper is to provide  a detailed description of the design and solutions adopted for the technical challenges arising during the development of ESASky, as well as for the implementation of its most relevant features.

The structure of the paper is as follows: section \ref{sec:hard} describes the overall hardware infrastructure of ESASky. The available interfaces are discussed in section \ref{sec:interface}, and the main data services and features are presented in sections \ref{sec:data services} and \ref{sec:features} respectively. Finally, the conclusions are given in Section \ref{sec:conclusions}.

\section{Hardware Infrastructure}\label{sec:hard}

\begin{figure}
	\centering
	\includegraphics[width=9cm]{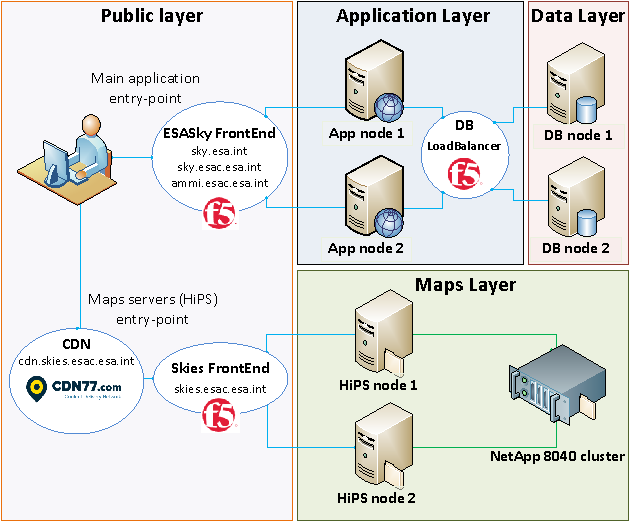}
	\caption{ESASky scalable hardware architecture.}
    \label{fig:ESASkyArchitecture}
\end{figure}

One of the main drivers for the hardware design was to provide the best possible user experience in terms of fast response time and high availability. Thus, the architecture was designed to be modularised and easily scalable from the beginning of the project.

ESASky infrastructure can be broken down into three distinct main layers, each  corresponding to a group of servers as described in figure \ref{fig:ESASkyArchitecture}. The application layer is composed of two small virtual machines hosting both front-end and back-end, which are Java web applications deployed into an Apache Tomcat 8 application server. The data layer includes two powerful physical servers hosting the databases. Finally, the maps layer contains two HTTP servers serving the HiPS all-sky maps.

On top of each group there is a dedicated load balancer in charge of distributing the user's load over the servers behind it. The total absence of the concept of a user session in ESASky makes it relatively easy to scale down the load by cloning and adding one or more servers to the proper load balancer.

Servers and load balancers are physically hosted at ESAC, Madrid. Therefore the responsiveness of the application, in terms of visualising and retrieving data, decreases with distance to the physical servers due to network latency. To solve this issue, ESASky uses an external Content Delivery Network (CDN) to enable fast visualisation of the all-sky maps from anywhere in the world.

\subsection{Public Layer}
The public layer consists of a CDN entry point for the static contents included in the web GUI, and two public load balancers, "ESASky FrontEnd" and "Skies FrontEnd" (see figure \ref{fig:ESASkyArchitecture}), in charge of distributing the user requests among the application and maps servers by following a round robin policy.
 
The CDN allows static contents of a web site to be added into cache nodes spread around the world. By uploading contents into the CDN storage node, they are automatically synchronised into the network of CDN cache nodes on demand.

Currently, the ESASky CDN service is configured to have a single storage server located in Prague, Czech Republic. From there, ESASky contents are cached in other parts of the world depending on the user demand. Therefore, users sitting further from the central storage node may experience lower start-up if the data have still not been replicated in the internal CDN network of cache nodes.

ESASky relies on the CDN for serving the static content of the application, such as CSS files, Javascript libraries, icons, JSON configuration files, HiPS and dictionaries. Thanks to the use of the CDN, the average loading time of the application has decreased by a factor of 10 (from 30 seconds down to 3 seconds on average).

The main drawback is a more complex infrastructure for the deployment of the application since part of its contents is cached in the CDN nodes and they need to be refreshed on demand.

\subsection{Application Layer}
The application layer is composed of two different components, front-end and back-end. Both of them share a common instance of a Tomcat application server.

The front-end is developed in Java using Google Web Toolkit\footnote{http://www.gwtproject.org}. This front-end aims to be an easy-to-use portal for accessing data from an ever increasing list of astronomy missions without loosing the specific scientific background behind them.  

The back-end is a Java web application integrating a TAP+ server (see section \ref{sec:interface}), reachable via the URL http://sky.esa.int/esasky-tap/ in a synchronous mode.

\subsection{Data Layer}\label{sec:data layer}

The data layer is composed of two identical physical servers with 64 CPUs Intel(R) Xeon(R), 512 GB of RAM, and 3TB of solid state disk. Each server has been equipped with a PostgreSQL v9.5.2, Q3C\footnote{https://github.com/segasai/q3c} and pgSphere\footnote{http://pgsphere.github.io/} extensions installed. One of the main constraints is the full-time availability of the data by avoiding downtime periods or interruption of services when updating the data provided. This has been achieved by identifying two different groups of dataset types that are mapped into the database structure:
%
%

\begin{itemize}
\item Static physical tables for datasets that do not require update, such as catalogues and data coming from legacy missions.
\item Materialized views created on datasets coming from missions still in operation, whose data is updated regularly. Foreign tables are used to link the database to other archive databases and then the data is materialized locally.
\end{itemize}

The benefit of using a materialized view is the possibility to update its content without causing any downtime or unavailability of services by refreshing concurrently. Another benefit is that materializing the data from the foreign tables makes the access much faster as it is accessed locally (including indexes) avoiding any latency issues. The drawback is that the concurrent approach makes the update slower than a simple refresh which is more appropriate when the table is bigger. The maintenance of the database also requires more work due to the increasing space used. Alternatives to the concurrent refresh are being investigated at the moment.

The second constraint on the data layer lies in the responsiveness of the whole system where the database itself could become its bottleneck. A study was carried out to select the proper indexing on the tables since most of ESASky database queries rely on geometrical searches, very expensive in terms of CPU cost. To tackle this constraint, two geometrical index approaches were tested with pgSphere and Q3C extensions, and a final solution was found depending on the types of datasets involved:

\begin{itemize}

\item \textit{Catalogue Datasets:}

This type of dataset is represented in ESASky by their catalogue coordinates ($\alpha,\delta$) as a point in the sphere. Q3C was found as the best performing index for the following type of queries:

\begin{equation}\label{q3c query}
\rm{source}(\alpha,\delta)\in \rm{FoV}
\end{equation}
where FoV is the field of view of interest represented as an array of points. 

\item \textit{Observation Datasets:}

The image and spectra observations are geometrically represented in ESASky as the so-called "footprints", which describe the particular instrument or detector sensitive exposure area in the sky. In this case, pgSphere index was selected for the following type of queries:
\begin{equation}
\rm{footprint_{obs}} \cap FoV \ne \emptyset
\end{equation}
\begin{equation}
\rm{footprint_{obs}} \subseteq FoV \ne \emptyset 
\end{equation}
\begin{equation}
\rm{footprint_{obs}} \supseteq FoV \ne \emptyset
\end{equation}
where footprint$_{\rm{obs}}$ is a \textit{spoly} pgSphere data type representing the sensitive exposure area of an observation and FoV is the Field of View in the application.

\end{itemize}

\subsection{Maps Layer}

The Maps layer is the physical storage for the HiPS served, composed of two small virtual machines (2 CPUs, 4GB RAM). Each of them provides HiPS through a simple Apache httpd 2.x service through a round robin load balancer. ESASky uses these servers to synchronise HiPS with the CDN. This layer is also becoming an official HiPS mirror as part of the HiPS network.

\subsection{ESAC Grid}\label{subsec:GRID}
The ESAC Grid infrastructure serves as a cluster computing environment for the processing of heavy tasks. It is based on the Univa Grid Engine\footnote{http://www.univa.com} which efficiently schedules multiple jobs into a common set of computing resources available for all the missions at ESAC. The cluster includes a submission host for the user login and preparation of the tasks, an admin host which receives, queues and performs the scheduling according to priorities and load distribution and a set of 46 execution nodes. The tasks are finally run in parallel as jobs on the execution nodes and the output data written into a common NFS data volume accessible by the rest of the infrastructure elements.

As a reference of the usage, ESASky ran more than 500,000 jobs, consuming 41.4 CPU years during 2017. The cluster comprised of a total of 1376 compute units (CPU slots) and an aggregate of 3.5 TB of RAM. 

\section{Interfaces}\label{sec:interface}
In this section the main interfaces of the application are described, and are classified by their access point.

\subsection{Back-end}\label{subsec:interface-be}
The most important entry-points to the services provided are:

\begin{itemize}

\item \textit{TAP+ server:}
This is an extension of the IVOA Table Access Protocol (TAP; \citealt{TAP}), which specifies how to interact with data by using an SQL-like language adapted for astronomy called Astronomical Data Query Language (ADQL; \citealt{ADQL}). It was first developed for the Gaia archive \citep{Gaia} to extend the original functionalities. 

\item \textit{Search:} 
This service is in charge of resolving the input provided by the user through the web GUI. The back-end contacts SIMBAD\footnote{http://simbad.u-strasbg.fr/simbad/} for astronomical objects and SsODNet\footnote{http://vo.imcce.fr/webservices/ssodnet/} for solar system objects to retrieve the position coordinates and galdim details when applicable. These details are used in the web GUI for the visualisation and the computation of the best FoV. Moreover, there is a quartz job scheduler\footnote{http://www.quartz-scheduler.org/} service checking regularly the availability of the official Strasbourg and Harvard SIMBAD mirrors.
 
\item \textit{Science-Ready Data Download:}
This is a synchronous download service acting as proxy to the archives where the data are located. The links to the original data products are stored into the application local database. 

\end{itemize}

\subsection{Web GUI}
\begin{figure*}
	\centering
	\includegraphics[width=16cm]{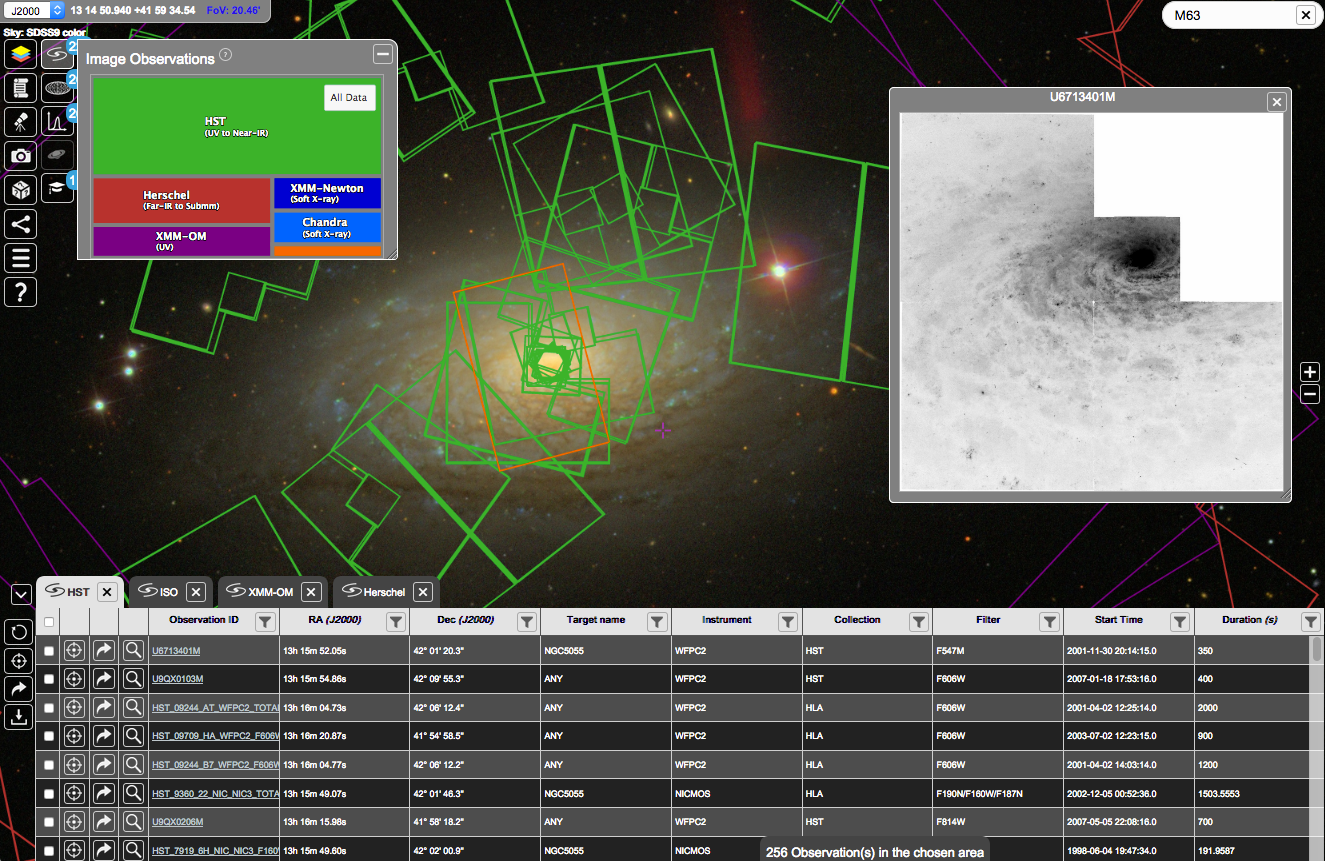}
	\caption{A view of ESASky showing the treemap, footprints, data table and an HST image postcard.}
    \label{fig:ESASkyWithDataPanel}
\end{figure*}

ESASky utilizes a HiPS viewer (AladinLite\footnote{http://aladin.u-strasbg.fr/AladinLite}) to let users navigate through the whole sky. To explore data from a certain astronomical object or region, one can navigate to the desired object in the sky or type the object's name into the search box. The Web GUI will then show the available data by looking at the total count number next to each data type (observations, catalogues, spectra), as shown in figure \ref{fig:ESASkyWithDataPanel} on the top left corner. This number is computed by a dedicated Fast-Count algorithm implemented at the database level (see section \ref{fast-count} for more details).

To display the large amount of data available in a user friendly way, it was decided to integrate the flowchart Highcharts\footnote{https://www.highcharts.com} JavaScript library to visually represent each dataset as a branch (box) in a logarithmic treemap (see Fig. \ref{fig:ESASkyWithDataPanel}). This implies that each box area is proportional to the relative amount of data available in the current FoV with respect to the total count number. Users can click on a box to retrieve all available metadata and footprints associated to the chosen region. 

Metadata are displayed in a table format within the main data panel (see figure \ref{fig:ESASkyWithDataPanel}). The direct download of the data products is performed through this data panel calling the dedicated Science-Ready Data Download back-end interface described in section \ref{subsec:interface-be}.

The web GUI gives users intuitive access to data from an ever increasing list of astronomy missions without the need of writing a single query. This interface can also communicate with other VO tools using the IVOA standard SAMP \citep{SAMP}. This enables even greater possibilities for science exploitation by using other powerful programs such as DS9, Topcat, and Aladin Desktop.

The web GUI provides additional interfaces:
\begin{itemize}

 \item \textit{Targetlist file upload:}
This allows an input file with a list of target names or coordinates to be uploaded through the web GUI. Once parsed and validated, it becomes available as a plain list on a specific target list panel to easily navigate to each input provided, allowing visual cross-matching against available datasets. 

\item \textit{Publication Retrieval:}
This interface provides the metadata related to a requested publication, relying on the dedicated SIMBAD public TAP service. This service is invoked by passing either a bibliographical code or an author as parameters within the URL query string. 
\end{itemize}

\subsection{Mobile}
Support to mobile browsers was added in the latest release of ESASky (version 2.1), which gives users access to the power of ESASky while not on their workstation. Further usability improvements, as well as a mobile web-based app, are planned to be released in future versions.

\subsection{AstroQuery}

For download and processing huge amounts of data, a Python command line interface is available. The interface is created as a module of Astroquery\footnote{http://www.astropy.org/astroquery} \citep{Astroq}, which is an astropy\footnote{http://www.astropy.org} affiliated package (\citealt{Astropy1}, \citealt{Astropy2}). This gives users direct access to the data hosted in ESASky either from astropy repository tools or from their own command-line scripts.

\section{Data Services}\label{sec:data services}

The core data service is providing access to the high-level metadata and data products hosted in the ESDC astronomy archives. In order to avoid overloading the web GUI browser and keeping all related scientific information, two different approaches are in place depending on the nature of the data itself.

\begin{itemize}

\item \textit{Catalogue Metadata}
A predefined limit of 3000 catalogue sources is set for queries generated by the web GUI. The results of these queries are sorted over a predefined column for the selection of the first N rows, therefore the sorting criteria is specific for each catalogue.

\item \textit{Observation Metadata} 

Since there is no natural geometrical ordering of the observation metadata, a HEALPix Multi-Order Coverage map (MOC; \citealt{MOC}) is returned when the fixed limit of 3000 footprints is reached. Each dataset has a related MOC table generated in-house to represent visually its coverage in the sky. This MOC table is stored as a sequence of HEALPix \citep{KG:2005} tiles parsed as \textit{spoly} pgSphere objects in the database.

The decision to perform the query against either the dataset table or the MOC table relies on a Fast-Count strategy (see section \ref{fast-count}) depending on the total number of footprints returned within a given FoV. 

\end{itemize}

\begin{figure*}[!tbp]
  \centering
  \begin{minipage}[b]{0.4\textwidth}\label{fig:1sso-xmatch}
    \includegraphics[width=\textwidth]{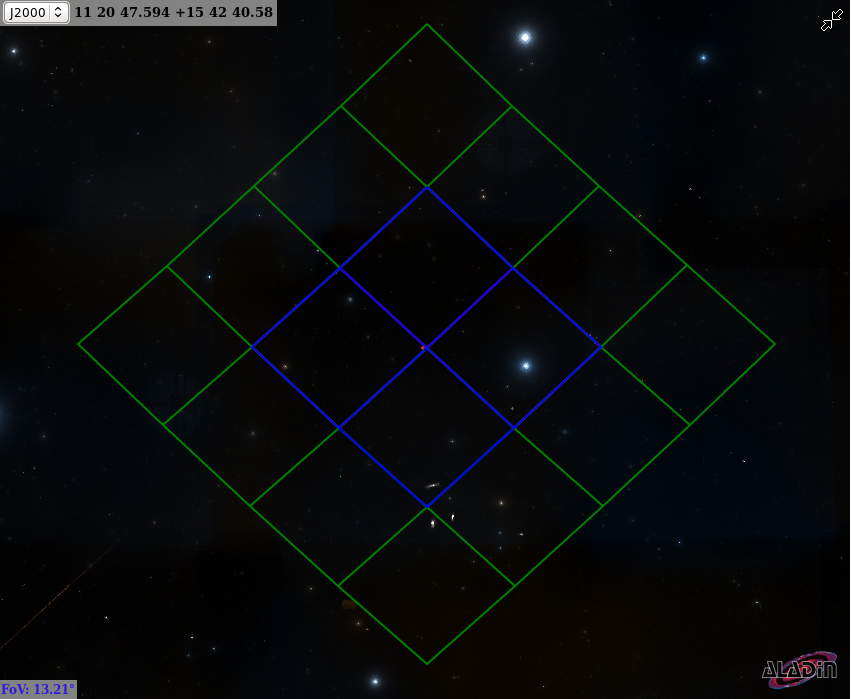}  
  \end{minipage}
  \hspace{1em}
  \begin{minipage}[b]{0.4\textwidth}\label{fig:2sso-xmatch}
    \includegraphics[width=\textwidth]{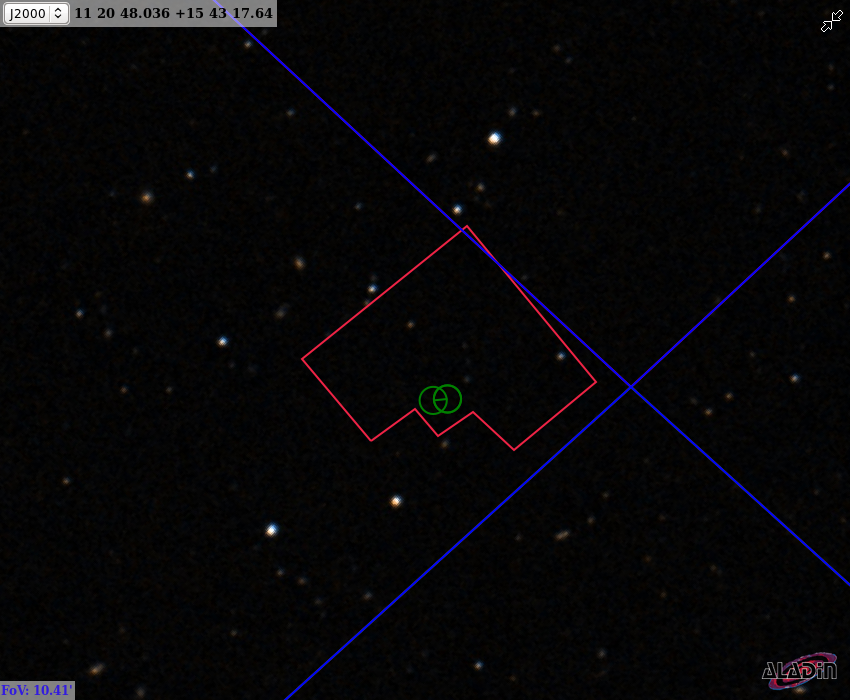}
  \end{minipage}
\caption{Example of the two-step cross-match process followed by the comet 67P against an HST observation "\textit{HST\_09713\_02\_WFPC2\_F555W\_WF}". a) On the left, the first cross-match performed by comparison of the HEALPix pixels representing the maximum position uncertainty of 67P in $\Delta$t = 10 days (green tiles), and the HEALpix tiles from the observation footprint (blue tiles). b) On the right, a zoom-in of the area where the second cross-match step is displayed. Green circles represent the SSO recomputed position and uncertainties (exposure start/end times). In red is the HST observation footprint.}
\label{fig:xmatch}
\end{figure*}

\subsection{Fast-Count}\label{fast-count}

The Fast-Count service provides a fast count of all available datasets in a particular area of the sky before retrieving the actual metadata. This service was developed to deal with the long response time that usually arises from counting over huge database tables in a large FoV scenario. It is used in two contexts:
\begin{itemize}
\item To discern between a MOC or real observation metadata retrieval. 
\item To update the approximate number of data available in the current FoV.
\end{itemize}

The logic behind the fast count is based on a HEALPix tessellation of the sky. Depending on the sky area, the service implements a decision logic to return either an estimated or a real count of available metadata within the current FoV. The entry point is an SQL function that accepts as parameters the FoV and its centre. This function queries a structure in the database representing four density distribution maps, each one related to a tessellation level of the sky (nsides 32, 64, 128 and 256). All data collections are mapped against these four different density distribution maps, so a precomputed estimation is stored for the number of results per data collection and per HEALPix pixel.

In addition to this core functionality, additional services have already been implemented in an effort to allow a larger exploitation of the data available. 

\subsection{Solar System Objects}

Allowing the Solar System community fast and easy access to the astronomical data archives is a long-standing issue. Moreover, the everyday increasing amount of archival data coming from a variety of facilities, both from ground-based telescopes and space missions, leads to the need for single points of entry for exploration purposes. 

Efforts to tackle this issue are already in place, such as the CADC Solar System Object Image Search\footnote{ttp://www.cadc-ccda.hia-iha.nrc-cnrc.gc.ca/en/ssois/}, plus a number of ephemeris services like NASA-JPL Horizons\footnote{https://ssd.jpl.nasa.gov/horizons.cgi}, IMCCE Miriade\footnote{http://vo.imcce.fr/webservices/miriade/} or the Minor Planet \& Comet Ephemeris Service\footnote{https://www.minorplanetcenter.net/iau/mpc.html}. 

Within this context, a first integration of the search mechanism for Solar System Objects (SSOs) was developed for ESASky. Based on the IMCCE Eproc software \citep{Berthier1998} for  ephemeris computation,  it  allows fast  discovery of photometry  observations from a subset of ESA missions that potentially contain those objects within their FoV. 

In this first integration the user is able to input a target name, resolved against the SsODNet service\footnote{http://vo.imcce.fr/webservices/ssodnet/}, and retrieve on-the-fly the results for all the observations matching the input provided. The results are precomputed over a geometrical cross-match between the observation footprints and the ephemeris of the SSOs within the exposure time frame of the observations.

The processing pipeline was developed in the Java programming language and runs in the ESAC Grid infrastructure described in section \ref{subsec:GRID}. The workflow can be described in three separate steps:

\begin{itemize}

\item \textit{Ephemeris Computation}: a first computation for each object is performed and stored for a defined time-step of 10 days from each of the satellites reference-frame (XMM-Newton, Herschel and the Hubble Space Telescope; HST).

\item \textit{Geometrical Cross-Match}: Two consecutive cross-matches are performed in order to speed the global computational time. 

The first cross-match serves as a fast selection of possible candidates decreasing the number of geometrical cross-matches needed in the second step. In a first step, the position and uncertainty of each SSO coming from the previous step ($\Delta$t = 10 days) is cross-matched against the selected datasets imaging footprints. This cross-match is done based on HEALPix indexes (see left image of figure \ref{fig:xmatch}). The selection of the HEALPix order is computed based on the distance to the object and its proper motion. 

The output list of candidate observations per SSO undergo then a new precise geometrical cross-match where the position of the SSO is re-computed using the start time and duration of the observation and the cross-match is performed against the observation footprint (see right image of figure \ref{fig:xmatch}).

\item \textit{Ingestion in the Database}: Positive cross-matches are stored into a dedicated schema in the database with all the derived parameters calculated in previous steps (position, position uncertainty, apparent magnitude, distance, proper motion).
\end{itemize}

The input tables with the orbital parameters used for the computation of the ephemeris are: the Lowell Observatory Asteroid Orbital Parameters table \footnote{http://asteroid.lowell.edu}, and the orbital elements of comets computed at the IMCCE\footnote{http://www.imcce.fr/en/ephemerides/donnees/comets/index.html}.


\section{Features}\label{sec:features}
This section covers a number of features developed with the purpose of providing extra functionality not linked to the existing archival data sitting at the ESDC repositories described in the previous section.

\subsection{Mission planning tool}

\begin{figure}[!ht]
	\centering
	\includegraphics[width=6cm]{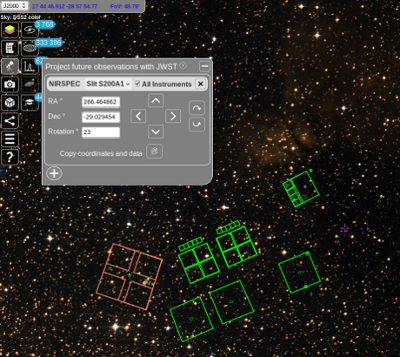}
	\caption{A view the Mission Planning Tool with the JWST Focal Plane Assembly. In red, the NIRSPEC instrument centred on the selected detector.}
   \label{fig:jwst-planning-tool}
\end{figure}

Thanks to its visual approach to data, ESASky renders itself as a perfect holder for the so-called "Mission Planning Tool". This feature, shown in figure \ref{fig:jwst-planning-tool}, allows users to project a given instrument onto the sky and change the position angle and coordinates not only of the whole Focal Plane Assembly (FPA) but also of its individual detectors. This feature offers multiple advantages to the final user in terms of usability and exploitation of the existing archival data for comparison purposes.

It is a standalone Java library, accepting as input a given instrument or particular detector and position, and returning a map with the projected sky coordinates for the entire FPA centred on the input given. The telescope reference frame V2V3 coordinates coming from the Science Instrument Aperture File (SIAF) provided by the mission are transformed to their sky equivalents ($\alpha$,$\delta$) using a sequence of five rotation matrices, where $r_{0}$ is the local roll, defined as the position angle (measured from North over East) of the V3 axis.
\begin{equation}
\omega(\alpha,\delta)=M(v2_{0},v3_{0},\alpha_{0},\delta_{0},r_{0})\omegaˆ{'}(v2,v3)
\end{equation}

The planning tool is currently available for the James Webb Space Telescope (JWST), and it is planned to incorporate other ESA missions soon.

\subsection{Publications}

\begin{figure*}
	\centering
	\includegraphics[width=16cm]{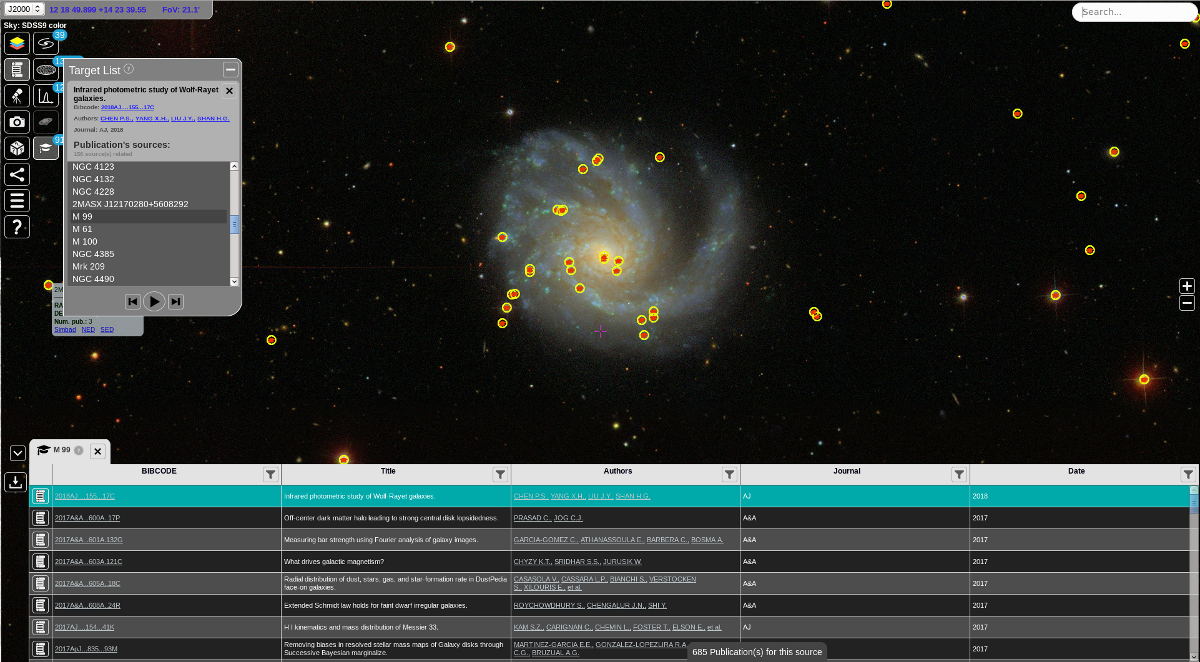}
	\caption{A view of ESASky showing the publications related to Messier 99.}
    \label{fig:publications}
\end{figure*}

The integration of ADS within ESASky was released in the latest version (v2.1) in February 2018. This feature displays all publications related to an astronomical source following the same client logic implemented in the rest of the application (see figure \ref{fig:publications}). Thus, a dedicated button was placed in the main control tool bar of the application to enable users to visualise all sources with at least one publication available within the FoV. On selection of a particular source, the detailed list of all related publications will be shown in the main data panel.

This feature depends on the SIMBAD TAP service from CDS, since it queries SIMBAD dedicated publication tables to retrieve the number of publications per source, the list of sources linked to a publication and its details.

To avoid performance issues a set of limits were established on the number of sources displayed in the sky (500) and the number of rows in the data panel (3000). The latter was selected after an extensive analysis of the distribution count of sources per publication. It was found that at least 99,9\% of publications had less than 3000 sources, so it was decided that this was a good limit.

\section{Conclusions}\label{sec:conclusions}  

ESASky provides a novel way to exploit the astronomical data sitting within ESAC archives and external partners from a visual perspective. Its design allows a fast and responsive web application and ensures the scalability of the services provided, taking advantage of the existing IVOA protocols and extensions developed within the ESDC group.  

The main driver of the application is serving high-level data products and metadata from the astronomy archives hosted at ESAC and by external partners. The internal design of the application allows the integration of new astronomical data by simply intervening at the database level. In addition, the data services described in this paper provide extra entry-points to these metadata for new scientific exploitation of the data, for which new algorithms and solutions have been described. 

Future developments will include cut-out services and integration of a new time-domain context.

\section*{Acknowledgments}

If you have used ESASky for your research or work, please acknowledge the tool in your publications\footnote{https://www.cosmos.esa.int/web/esdc/esasky-credits}. This URL lists all the people and institutes that have contributed to its creation. This work also benefited from experience gained from projects supported by both ESA SCI-OO and SCI-OP research funding.


\end{document}